\documentclass[12pt, preprint2]{aastex}
\usepackage{geometry} 

\slugcomment{Accepted to AJ}

\shorttitle{Variability of Deneb} 
\shortauthors{Richardson et al.} 

\begin{document}


\title{A Five-year Spectroscopic and Photometric Campaign on the Prototypical $\alpha$ Cygni Variable and A-type Supergiant Star Deneb}
\author{N. D. Richardson\altaffilmark{1}, 
N. D. Morrison\altaffilmark{2},
E. E. Kryukova\altaffilmark{2},
S. J. Adelman\altaffilmark{3}}

\altaffiltext{1}{Center for High Angular Resolution Astronomy, 
Department of Physics and Astronomy, 
Georgia State University, P. O. Box 4106, Atlanta, GA  30302-4106; 
richardson@chara.gsu.edu} 
\altaffiltext{2}{Ritter Astrophysical Research Center, Department of Physics and Astronomy, University of Toledo, 2801 W. Bancroft, Toledo, OH 43606; nmorris@utnet.utoledo.edu, eallga@physics.utoledo.edu}
\altaffiltext{3}{Department of Physics, The Citadel, 171 Moultrie Street, Charleston, SC 29409; adelmans@citadel.edu}

\begin{abstract}

Deneb is often considered the prototypical A--type supergiant, and is one of the visually most luminous stars in the Galaxy. A--type supergiants are potential extragalactic distance indicators, but the variability of these stars needs to be better characterized before this technique can be considered reliable. We analyzed 339 high resolution echelle spectra of Deneb obtained over the five--year span of 1997 through 2001 as well as 370 Str\"omgren photometric measurements obtained during the same time frame. Our spectroscopic analysis included dynamical spectra of the H$\alpha$ profile, H$\alpha$ equivalent widths, and radial velocities measured from Si II $\lambda\lambda$ 6347, 6371. Time-series analysis reveals no obvious cyclic behavior that proceeds through multiple observing seasons, although we found a suspected 40 day period in two, non-consecutive observing seasons. Some correlations are found between photometric and radial velocity data sets, and suggest radial pulsations at two epochs. No correlation is found between the variability of the H$\alpha$ profiles and that of the radial velocities or the photometry. Lucy (1976) found evidence that Deneb was a long period single--lined spectroscopic binary star, but our data set shows no evidence for radial velocity variations caused by a binary companion.
\end{abstract}


\section{Introduction}

$\alpha$ Cygni variables are highly luminous OBA stars that exhibit low amplitude variations both in photometry ($\Delta V \lesssim 0.15$ mag; van Genderen et al.~1989) and velocity (e.g.~Abt 1957). These variations are common for massive stars and are observed in most hot supergiants as well as the over-luminous supergiants such as luminous blue variables. The variations have been observed in some main sequence massive stars, and the mechanism responsible for the variations has eluded theory and observers. Photometric and spectroscopic campaigns on multiple stars that span the upper H-R diagram and exhibit these variations will be crucial to understanding these variations and their source. 

Deneb ($\alpha$ Cygni, HR 7924, HD 197345) is an early--type A supergiant (spectral type A2Iae) and is the prototype for the $\alpha$--Cygni type variable stars. With a great apparent brightness and because it can be observed for most of the year at northern latitudes, Deneb can be easily studied spectroscopically with small or moderate aperture telescopes. Deneb has been one of the most studied A--type supergiants. Fath (1935) and Paddock (1935) made the first variability studies of this star, using photometry and radial velocities, respectively. Abt (1957) observed 9 supergiants, including Deneb, and showed that early and intermediate type supergiants exhibited oscillations of their atmospheres. Lucy's harmonic analysis of Deneb (1976) demonstrated that multiple oscillations were present in the atmosphere of the star. He also discovered what appeared to be a binary period in the radial velocities of Deneb. Parthasarathy and Lambert (1987) made 123 spectroscopic observations of Deneb in the near-infrared between 1977 and 1982 at McDonald Observatory. They found radial velocity variations similar to those observed by Paddock and analyzed by Lucy and claimed that the binary motion was recovered but presented no time-series analysis.

The underlying cause of the brightness and velocity variations of the $\alpha$ Cygni variables has been recently analyzed by a few authors. Saio et al.~(2006) found that long-period oscillations in HD 163899 could be understood as g-mode pulsations that are trapped in a convection zone positioned above the H-burning shell of the star. An investigation of Deneb's variability by Gautschy (2009) gave evidence that Deneb also has a sub-photospheric convection zone driving the variability. Cantiello et al.~(2009) showed how iron convection zones above the H-burning shell could be the cause of the microturbulence and clumping in the base of the wind of hot stars. The variability for the $\alpha$ Cygni variables decreases in amplitude and time scale
toward higher effective temperatures, and therefore these processes are easier to observe in the cooler A-type supergiants such as Deneb (e.g. Lefever et al.~2007).

Kaufer et al. (1996, 1997) examined the radial velocity variations and the H$\alpha$ profile variations of Deneb and five other supergiants. They found that the radial velocities were multiperiodic and showed different periods in different observing seasons. They also found that the equivalent width of H$\alpha$ exhibited multiperiodic variations and evanescent periodicities. In contrast, Morrison and Mulliss (1997) reported finding a much more active, and possibly cyclic, variability in the H$\alpha$ profile. Often there are ``absorption events" in which a secondary absorption component appears on the blue wing of the absorption component of the H$\alpha$\ profile. 

Literature on photometry of Deneb is remarkably sparse, mostly due to its great apparent brightness ($m_V = 1.25$), combined with the fact that there are no comparison stars of similar brightness in the vicinity of Deneb. Adelman and Albayrak (1997) used HIPPARCOS observations of Deneb, as well as other early A--type supergiants, to show that these objects are small or moderate amplitude variables (a few hundredths or a tenth of a magnitude in amplitude). The light curve was not complete enough to state more than there was a period on the order of about 2 weeks, but there were most likely other effects to take into account.

It has long been known (e.g., Abt and Golson 1966) that the strength of the H$\alpha$ emission in hot supergiants is correlated with the stellar luminosity. The potential of this fact for the use of these optically luminous stars as extragalactic distance indicators was greatly advanced when Kudritzki et al. (1999a,b) developed the wind momentum - luminosity relationship (WLR). However, only four A--type supergiants were in their sample, and the variability of most of these objects was neglected, with the exception of HD 92207 (A0Ia) which showed little variability in the momentum flow (${\dot M}v_\infty$) (Kudritzki et al.~1999b). The variability of these objects needs to be investigated further in order to quantify the likely error budget in the use of this relationship. With A--type supergiants being among the brightest stars in the optical, it is important to understand this relationship between momentum flow, as derived from an H$\alpha$ P Cygni profile, and the luminosity. 

The variability of the H$\alpha$ profile is an important diagnostic for hot supergiants in understanding the amount of clumping in the wind of these stars. If variations observed in H$\alpha$ profiles were found to be repeatable, they might be attributed to corotating interaction regions in the circumstellar matter (e.g. Cranmer \& Owocki 1996), which would constrain the rotation period of the star.

In this paper, we present numerous spectroscopic and photometric observations obtained during the calendar years 1997 through 2001. Such a large data set is important to understanding the complex observational variations in Deneb and other hot supergiant stars, especially in the context of the WLR. Further, variability studies in stars such as this could reveal the underlying physical mechanisms that cause the observed changes (e.g. Maeder 1980). For example, Gautschy (2009) demonstrated that some pulsational modes would be consistent with a deep convection zone for $\alpha$ Cygni variables.

\section{Observations}

339 high resolution ({\em R} = 26,000) red-yellow spectra of Deneb were obtained between 1997 March 20 and 2001 December 21 with the Ritter Observatory 1--meter telescope and fiber-fed \'echelle spectrograph. Details of the instrument are available in Morrison et al.~(1997)\footnote{\scriptsize A full listing of observations of Deneb at Ritter Observatory through 2006 is maintained online at http://astro1.panet.utoledo.edu/$\simeq$wwwphys/ritter/archive/Alpha-Cyg.html}. For the most part, one spectrum was obtained per night, although multiple exposures were obtained on some nights. Exposure times ranged from 60 to 1200 seconds, with most being about 300-400 seconds. The average signal-to-noise ratio was around 100 for the continuum near H$\alpha$. Portions of 9 orders of echelle data were recorded, and the orders including the H$\alpha$ profile and the Si II doublet at $\lambda\lambda$ 6347, 6371 are presented here. They were reduced using standard techniques for echelle spectroscopy via an IRAF\footnote{\scriptsize IRAF is distributed by the National Optical Astronomy Observatories, which are operated by the Association of Universities for Research in Astronomy, Inc., under contract with the National Science Foundation.} script.

370 differential Str\"{o}mgren photometric measurements were made with the Four College Automated Photoelectric Telescope (FCAPT) during the concurrent time period of 1997 October 29 to 2001 July 2. Non-variable check (ch) and comparison (c) stars were used. Each observation consisted of a measurement of the dark count, followed by the sequence sky-ch-c-v-c-v-c-v-c-ch-sky. The light from Deneb had to be passed through a 5 magnitude neutral density filter in order to avoid detector non-linearities. The light from the comparison and check stars was not passed through the filter. The estimated error in the differential photometry is about 0.004 magnitudes, as calculated from the standard deviation of the difference between check and comparison stars. After HD 197036 had been used as the comparison star for the first two years, with HD 198151 as the check star, it was discovered that HD 197036 is a small amplitude, long period variable star. Thus, it was replaced with HD 198151 for the remainder of these observations, and a new check star, HD 199311, was used. HD 197036 was not used for this analysis.

\section{Analysis}

Reductions beyond the observatory's pipeline script included removal of telluric lines, application of the heliocentric Doppler correction, and normalization to the continuum. Telluric lines were removed by means of the IRAF task {\tt telluric}. The reference spectra were artificial templates created by mapping telluric lines in the spectra of rapidly rotating hot stars taken with the same instrumentation. Each telluric line in the template was represented by a Gaussian of the appropriate equivalent width and full width at half maximum, superimposed on a unit continuum. 

When multiple exposures were taken in one night, the spectra were co-added before normalization, as spectral variability is not present on these time scales (e.g. Lucy 1976). From the continuum normalized spectra, dynamical spectra were created for the H$\alpha$ profile. When observations were not available on consecutive days, the spectra were linearly interpolated for the missing observations, provided the gap was less than 10 days. The net equivalent width of H$\alpha$ was measured for each of the continuum normalized spectra. The integration window was 22 \AA\ centered on the laboratory wavelength of H$\alpha$, and the estimated errors in the measurements are less than 5\%\ based on the techniques described by Chalabaev and Maillard (1983).  

Radial velocities were measured for the Si II $\lambda \lambda$ 6347, 6371 doublet by means of a Gaussian fitted to each line. A weak Fe II line was found to be blended into the blue wing of Si II $\lambda$ 6371, and was also fitted when determining the radial velocity of that line. When more than one observation was available for a single night, the radial velocity was measured for each observation; the variations were never more than 0.5 km s$^{-1}$ over the course of a night. Observation series of stable giant stars with this spectrograph and camera generally yield standard deviations of a single measurement on the order of 0.5 km s$^{-1}$. Typically, the radial velocities of the two Si II lines agreed with each other to within 1 km s$^{-1}$, but usually to within less than 0.5 km s$^{-1}$. 

The differential magnitudes were approximately converted to standard Str\"{o}mgren $uvby$ magnitudes, using values for the comparison stars from Hauck and Mermilliod (1998). The colors $u-b$ and $b-y$ and the Str\"{o}mgren indices $c_1$ and $m_1$ were calculated. Because no attempt was made to correct for any color dependence that may exist in the attenuation of the neutral density filter, and because multiple $uvby$ standard stars were not used, the colors we derived for Deneb are not precisely on the standard system. 

\section{Results and Time-Series Analysis}
\subsection{Dynamical Spectra of H$\alpha$}

Dynamical spectra were created for each observing season in order to explore the temporal behavior of the H$\alpha$ profile. Figure 1 shows the variations of Deneb during calendar year 1997. Several strong absorption events occur on the blue wing of the P Cygni profile at intervals of approximately 40 days. We define an absorption event for Deneb by the presence of a second absorption component in the blue wing of the principal absorption component. Usually this second absorption had a velocity blueshifted less than 100 km s$^{-1}$. The emission component remained fairly static, with the exception of two periods of strengthening. The absorption events we observed do not show a progression to the terminal velocity, as seen in discrete absorption components (DACs) of other hot, massive stars or in the high-velocity absorption events described by Kaufer et al. (1996) in hotter supergiants such as Rigel (B8 Ia). Aufdenberg et al. (2002) found that the terminal wind speed for Deneb is 225 km s$^{-1}$. Neither this observing season nor subsequent observing seasons show evidence that the H$\alpha$ absorption reaches the terminal wind speed. The line probably has a low optical depth at large distances from the star, due to insufficient population in the $n=2$ level.

\begin{figure}[htp]
   \centering
   \includegraphics[width=5in]{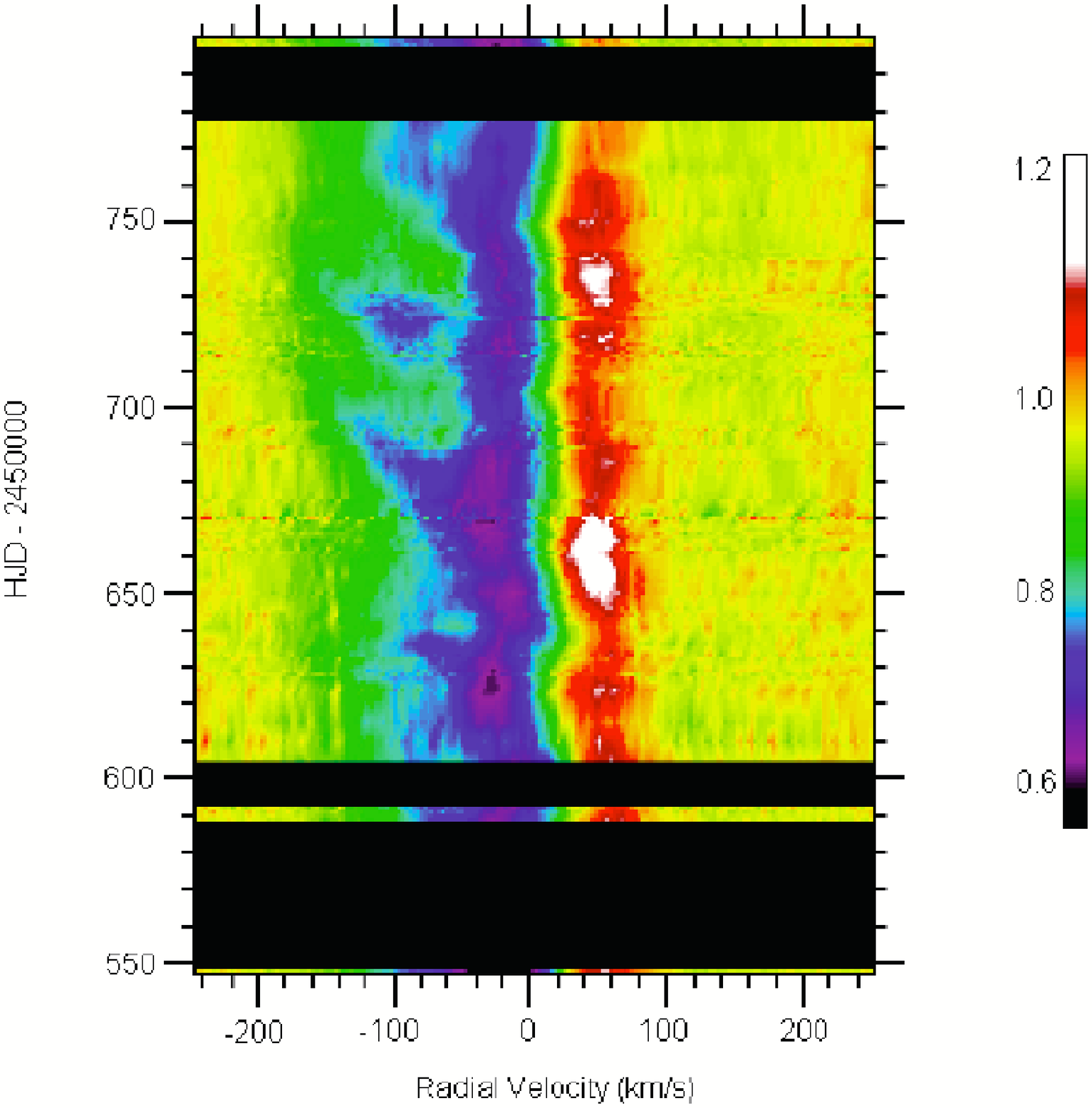} 
   \caption{\label{fig1} A dynamical spectrum of the H$\alpha$ profile for the year 1997. There are increases in the blue wing absorption (``absorption events") near HJD 2450635, 2450675, and 2450715.}
\end{figure}

\placefigure{fig1}

The cyclic-type behavior during the 1997 observing season was not repeated in the other observing seasons. The 1998 season (Figure 2) exhibits two absorption events, which are farther apart than the 40 day interval of the 1997 season. The emission feature is nearly static for that year of observations. The 1999 season (Figure 3) shows an exceptionally static H$\alpha$ profile for this star, with a possible beginning of an absorption event at the end of the observing season. One absorption event is observed in 2000 (Figure 4), but showed a more rapid onset and disappearance than the other observed events in our data set. The 2001 season (Figure 5) began with a very strong absorption event, and included one smaller event around HJD 2452075-2452100. The time span of September and October of 2001 (HJD 2452160-2452240) provided an interesting event that is discussed in Section 4.2. 

\begin{figure}[htp]
   \centering
   \includegraphics[width=5in]{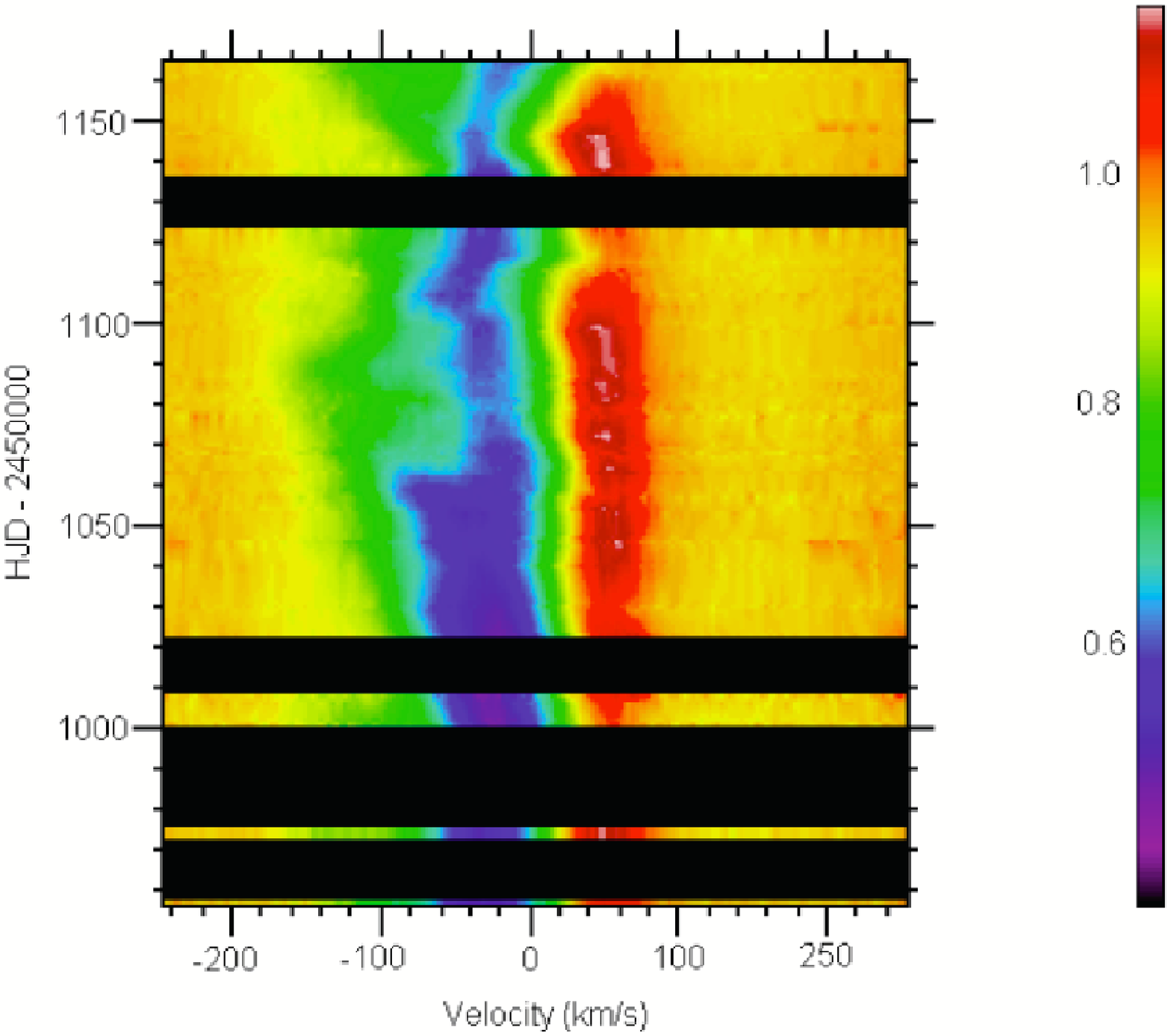} 
   \caption{\label{fig2} A dynamical spectrum of the H$\alpha$ profile for the year 1998.}
\end{figure}

\placefigure{fig2}

\begin{figure}[htp]
   \centering
   \includegraphics[width=5in]{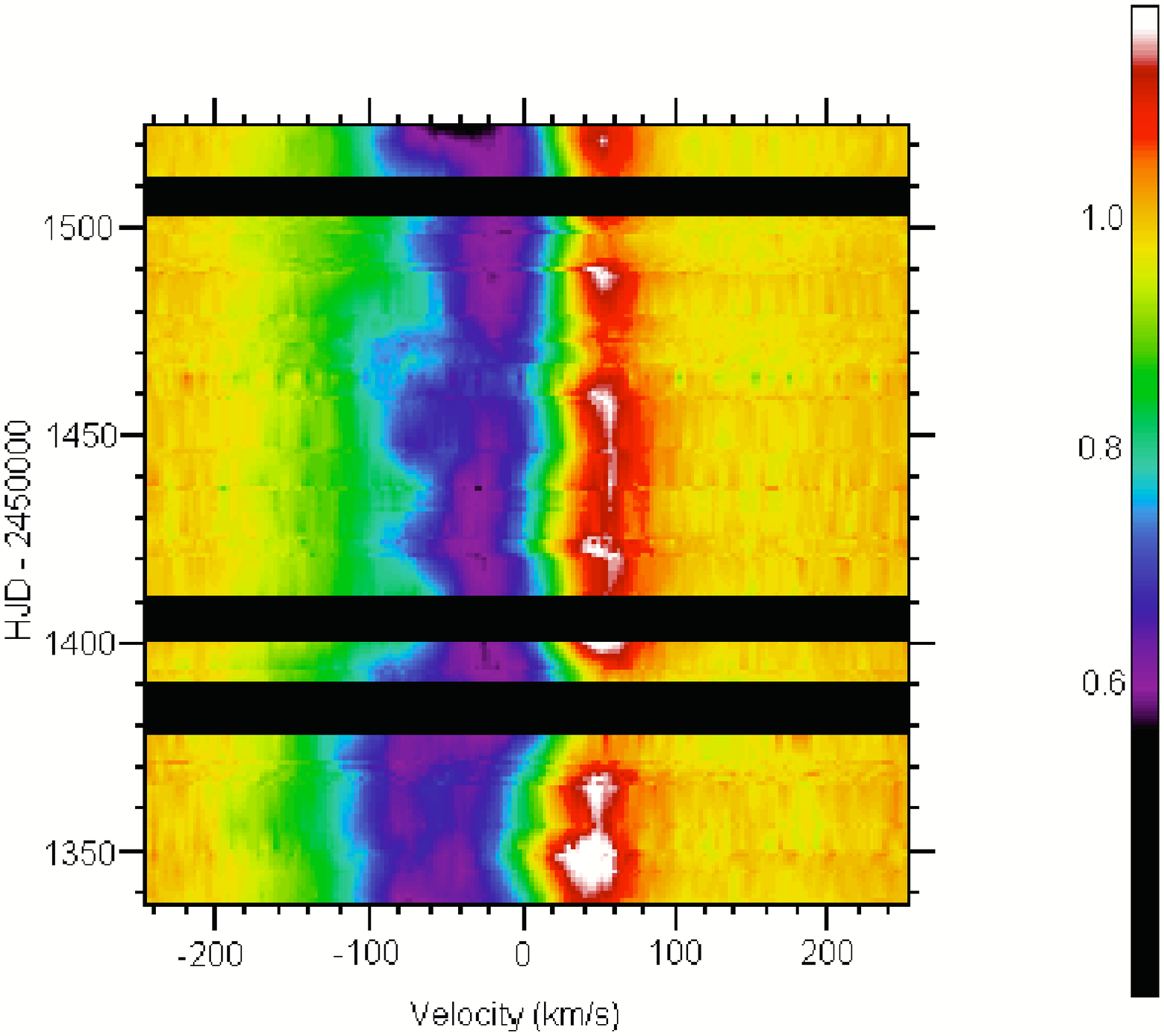} 
   \caption{\label{fig3} A dynamical spectrum of the H$\alpha$ profile for the year 1999.}
\end{figure}

\placefigure{fig3}

\begin{figure}[htp]
   \centering
   \includegraphics[width=5in]{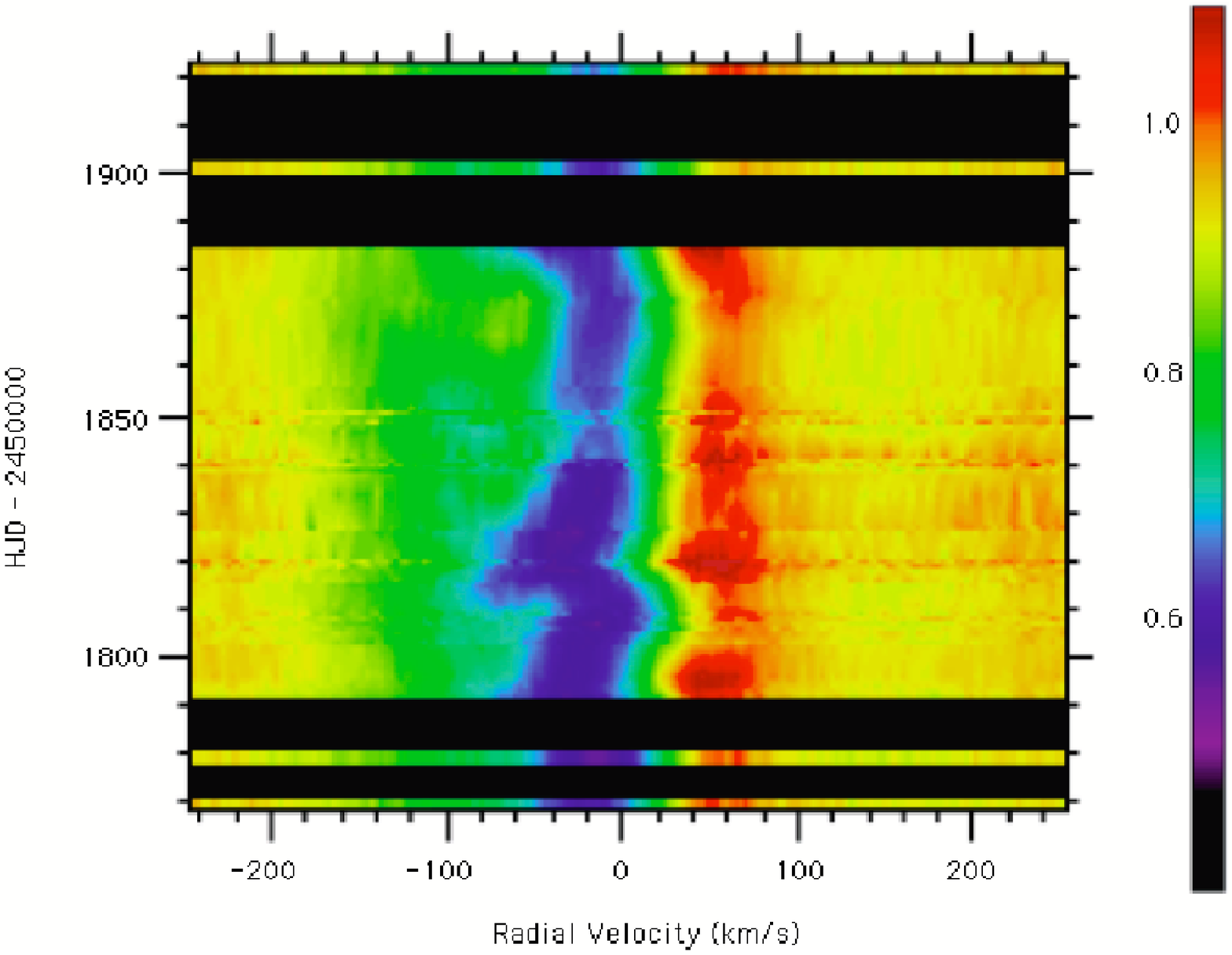} 
   \caption{\label{fig4} A dynamical spectrum of the H$\alpha$ profile for the year 2000.}
\end{figure}

\placefigure{fig4}

\begin{figure}[htp]
   \centering
   \includegraphics[width=5in]{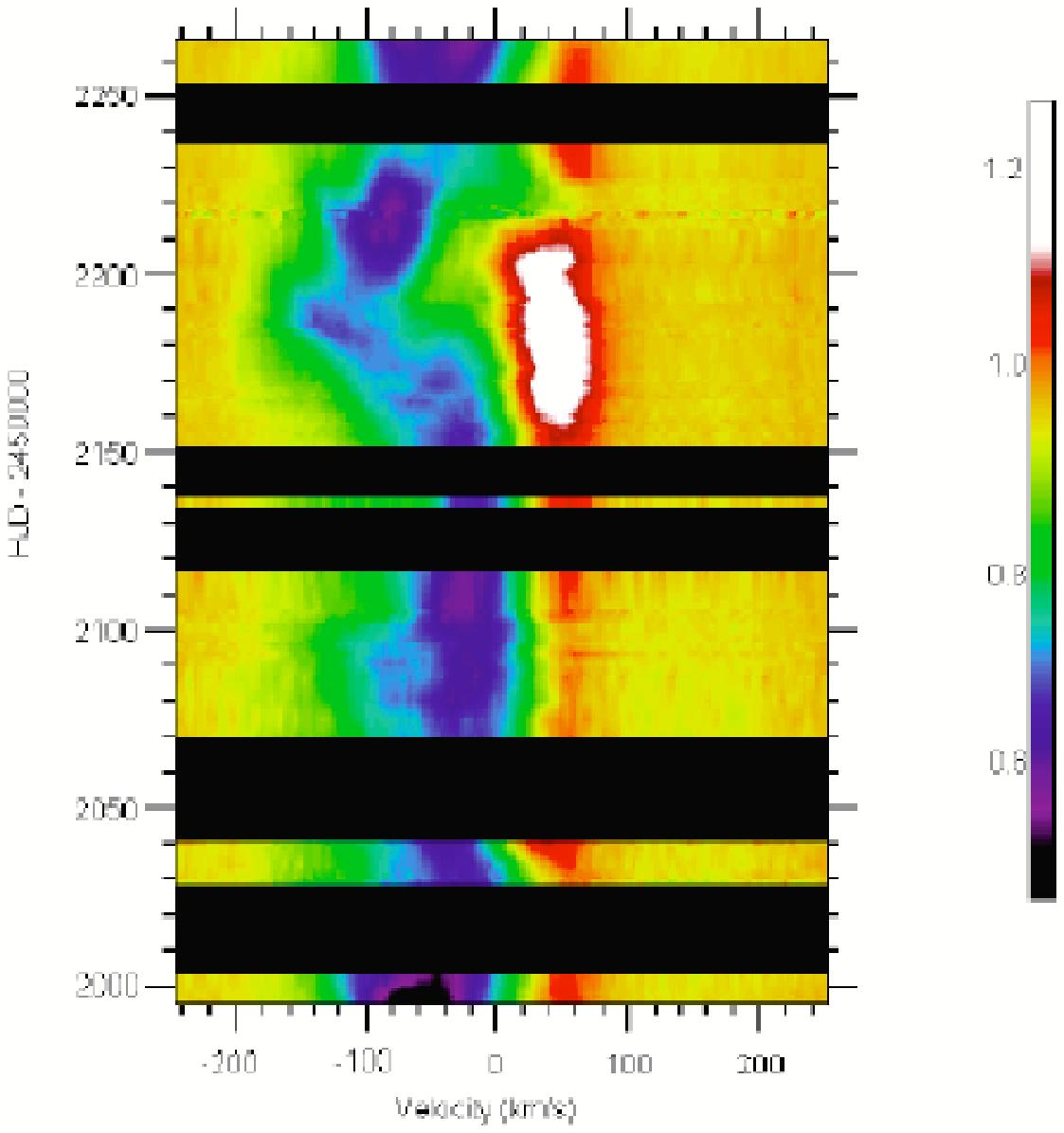} 
   \caption{\label{fig5} A dynamical spectrum of the H$\alpha$ profile for the year 2001.}
\end{figure}

\placefigure{fig5}

\subsection{The High-Velocity Absorption Event of 2001}
The remarkable event of late 2001 began with a quick onset of a strong emission component. The absorption component reached a larger negative velocity than had previously been observed by us or by Kaufer et al. (1996a,b, 1997). As the event subsided, we observed a quick and dramatic fading of the P Cygni emission. Such a small level of emission had not previously been observed in Deneb. Figure 6 illustrates the details of this event in line plots.

\begin{figure}[htp]
   \includegraphics[angle=90,width=6.5in]{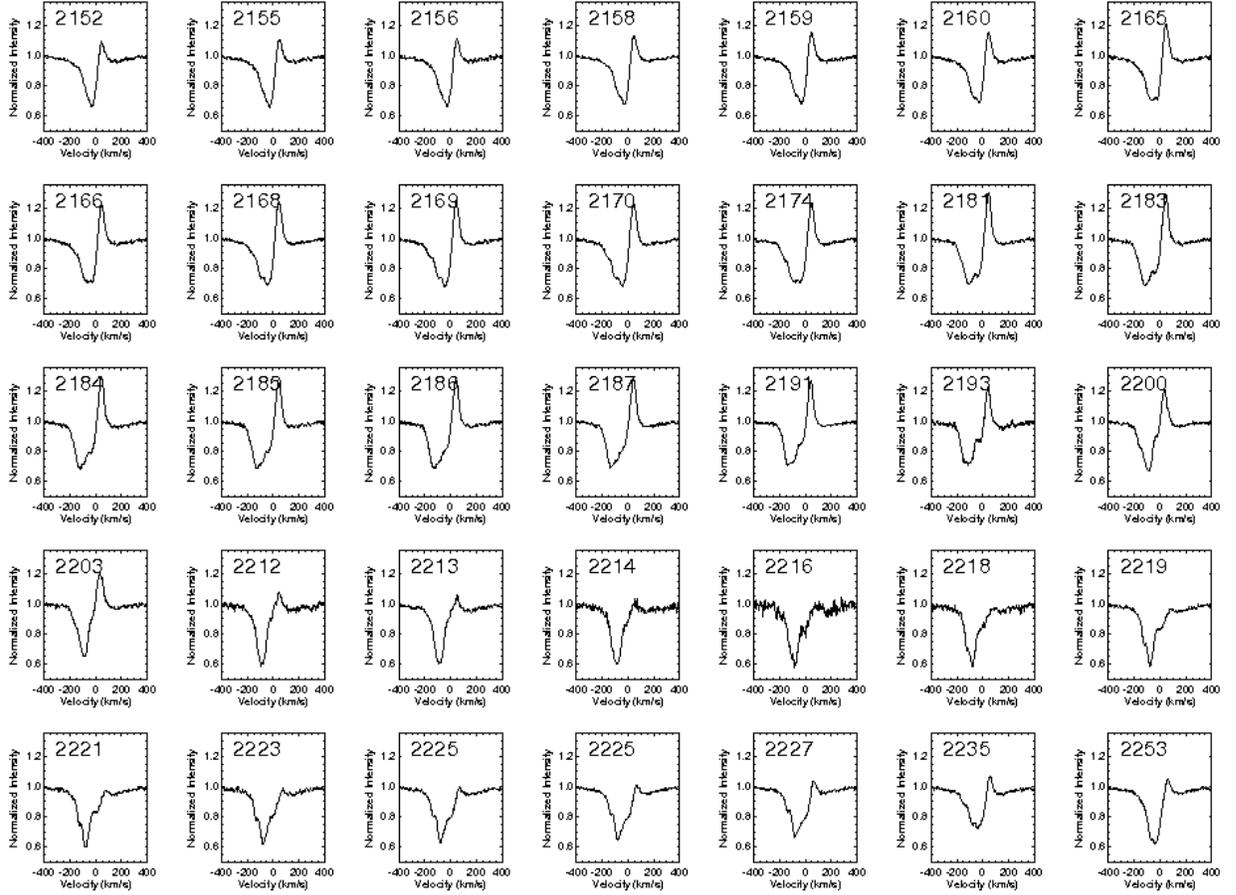} 
   \caption{\label{fig6} Line profiles of H$\alpha$ observed during the major 2001 absorption and emission event. Features to note are the development of deep secondary and tertiary absorption components, the increase in the H$\alpha$ emission and the near disappearence of the emission component beginning at HJD 2452214.5. All spectra are normalized to unity in the continuum and are plotted in the rest frame of H$\alpha$. The HJD - 2450000 is shown to the nearest integral day in each plot. }
\end{figure}

\placefigure{fig6}

The event also showed at least two additional absorption components at some epochs, and a dynamical spectrum created by subtracting the average line profile of the time period (Figure 7) reveals that this was actually two absorption events, spaced by the same period of 40 days observed with minor events in 1997. There is a sudden onset of a highly blueshifted ($ \sim -100$ km s$^{-1}$) absorption that increases in strength with time. About 40 days after the onset of this event, we see a second high velocity absorption event occur. As this second event reached its maximum absorption strength, we observed an onset of absorption on the red wing of the profile. This event fits the definition of a high-velocity absorption event, given in Kaufer et al.~(1996). These events are described as events, with a larger blue velocity and deeper absorption than normally observed, a similar morphology throughout the time development of the event, and no evidence of any spherical symmetry in the wind at these epochs. The high velocity absorption events are much more extreme than the ``normal" absorption events described in \S $4.1$.

\begin{figure}[htp]
   \centering
   \includegraphics[width=5in]{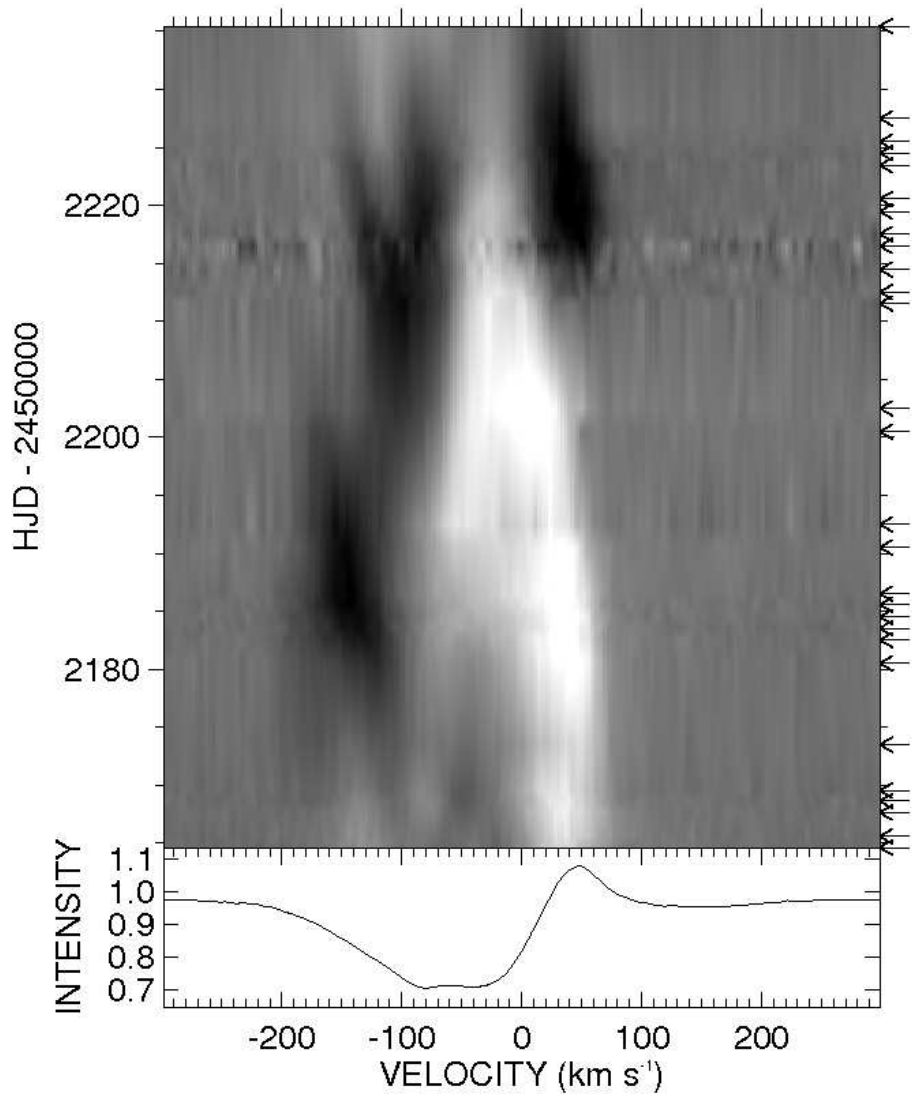} 
   \caption{\label{fig7} A dynamical spectrum of the H$\alpha$ profile during the 2001 high velocity absorption event. The average profile for this time period, plotted in the lower panel, has been subtracted. There is an onset of high velocity absorption, followed by a second absorption event $\sim$ 40 days later. The second portion of this event is followed by an onset of absorption on the red wing, causing the emission to appear the weakest that we observed during the 5 years of data presented in this study.}
\end{figure}

\placefigure{fig7}

\subsection{H$\alpha$ equivalent widths}

We measured the net equivalent widths of all our H$\alpha$ profiles (Table 1, available online). The errors in these net equivalent width measurements are typically 1\% of the measured value as determined using the methods of Chalabaev \& Maillard (1983). The H$\alpha$ profile shows a P Cygni type profile, so the net equivalent width does not provide a simple measure of all the variability. Using the IRAF task {\tt splot}, we measured the equivalent width of the emission component by trying to interpolate linearly across the base of the component. Unfortunately, this method is prone to personal error. From multiple measures from the same spectrum, we derived a typical error for the equivalent width of the H$\alpha$ emission component to be 5\% of the measured value. We then subtracted the emission component from the net equivalent width in order to obtain an absorption equivalent width.

We found that the net equivalent width had an average value of 1.318 \AA\, with a standard deviation of 0.197 \AA. The overall estimated strength of the absorption component(s) varied by slightly more than a factor of two ($W_{\lambda,{\rm H}\alpha,{\rm abs}}$=2.172 \AA~on HJD 2450999; $W_{\lambda,{\rm H}\alpha,{\rm abs}}$=1.067 \AA~on HJD 2450669). The absorption component was 1.616 \AA\ with a standard deviation of the mean of 0.209 \AA.

\subsection{Radial Velocities and Photometry}

In order to derive the systemic velocity of Deneb, we computed an unweighted average over all the observations of the Si II  6347, 6371 doublet, assuming no wind contamination of these lines. The result was -2.77 km s$^{-1}$, which is consistent with values derived from other studies (e.g.~Lucy 1976; Parthasarathy \& Lambert 1987). The radial velocity had a range of values of -10.8 to 9.1 km s$^{-1}$, consistent with the sum of the multiple pulsation mode amplitudes (10.44 km s$^{-1}$) derived by Lucy (1976).

The differential $uvby$ Str\"{o}mgren measurements (Table 2, available online) were averaged over all data to adopt the photometric parameters for Deneb given in Table 3. $u$, $v$, $b$, and $y$ had a range of 0.09, 0.14, 0.41, and 0.19 magnitudes respectively. The colors $u-b$ and $b-y$ had a range of 0.075 and 0.081 magnitudes respectively. The Str\"{o}mgren indices $c_1$ and $m_1$ were found to vary by 0.233 and 0.141 magnitudes, respectively.

\placetable{table3}

\subsection{Time-Series Analysis and the Binarity of Deneb}

In order to verify the reliability of various time-series analysis algorithms, their output was compared to that obtained by Lucy (1976) for the data collected by Paddock (1935). A Scargle periodogram analysis (Scargle 1982) of the data from Paddock (1935) only reproduced one aspect of the analysis of Lucy (1976), the long term variability that was suspected to be binary motion of Deneb. 

The confirmation of the binary nature of this prototypical supergiant would be useful for understanding the star, as well as its early evolution (including mass transfer) and estimating the current mass. We used the Scargle periodogram and the CLEAN algorithm (Roberts et al.~1987) to search our data set for the binary period, and found that the power spectrum reached a relative minimum in the period range that was suspected. As Lucy (1976) could not add other data to the data of Paddock to better constrain the orbit (Section III of his paper), and our data do not suggest any binary motion, the binary hypothesis for Deneb is not supported.

A method that is more robust for datasets containing multiple periodicities, the CLEAN algorithm was used to search for periodicities in the data sets. An analysis of the radial velocity data from Paddock (1935) produced similar results to that of Lucy (1976). The results are shown in Figure 8, where we plot the frequencies and their relative strengths found for each observing season for net H$\alpha$ equivalent width, Si II radial velocities, and Str\"{o}mgren $y$. The time-series analysis was performed on the entire data set, but because of the changing characteristics of the star, no global properties were found with a high probability of being real. If the same period were found for all three data sets, there would be some evidence for a photospheric-wind connection, but our analysis showed no evidence of this behavior. Similarly, if there were evidence for similar periods in just the photometry and radial velocities, that would provide information about pulsational behavior of the photosphere of Deneb.

\begin{figure}[htp]
   \centering
   \includegraphics[angle=90,width=6in]{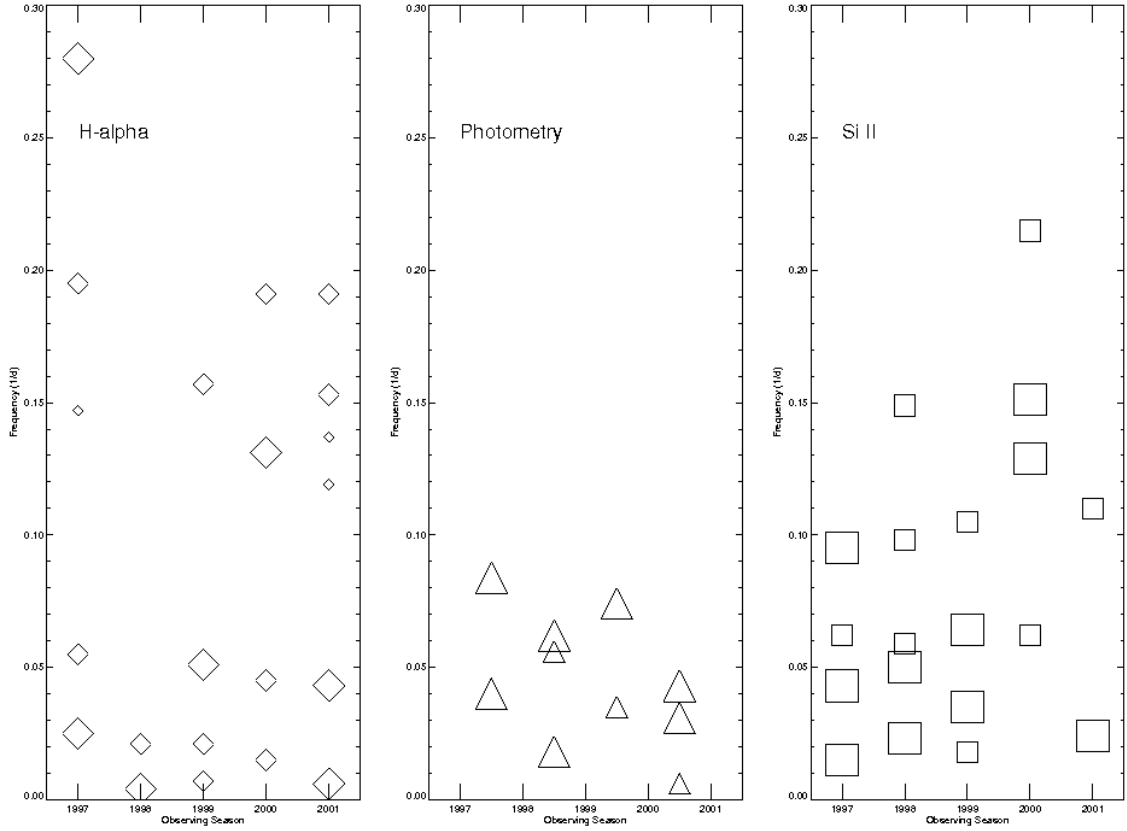} 
   \caption{\label{fig8} Plots of the frequencies derived in the CLEAN time series analysis. The symbol size is proportional to the relative strength of the derived frequency. The photometry produced similar results for all bandpasses, but only the $y$ filter is plotted.}
\end{figure}

\placefigure{fig8}

\subsection{Radial Pulsations}

Our analysis did find some periods that were the same for the radial velocities and for the photometry during 1998 and 1999. The phased (to HJD 2450000.0) data were fitted (minimum $\chi^2$) with a sine curve of the form 
\begin{equation}
N=A\times\sin(2\pi(p-x_c))+c
\end{equation}
where $N$ is the parameter we are fitting (radial velocity or magnitude), $p$ is the arbitrary phase we are solving for, $A$ is the amplitude, $x_c$ is a phasing offset, and $c$ is an offset for the systemic velocity or magnitude. The period was not allowed to vary during these fits. The parameters for our photometric and radial velocity fits are given in Table 4 for the two epochs of radial pulsations and the variations are shown graphically in Figure 9. The offsets in phase between the photometry maximum and radial velocity maximum for the two periods with the highest probability of radial pulsations (epochs with strong signals that match with both the photometry and radial velocity data; the fall of 1998 ($P=17.8$d) and the fall of 1999 ($P=13.4$d)) are roughly 0.25$P$, which strongly supports the hypothesis that Deneb was experiencing radial pulsations at these epochs.

\placetable{table4}

\begin{figure}[htp]
   \centering
   \includegraphics[angle=90,width=6in]{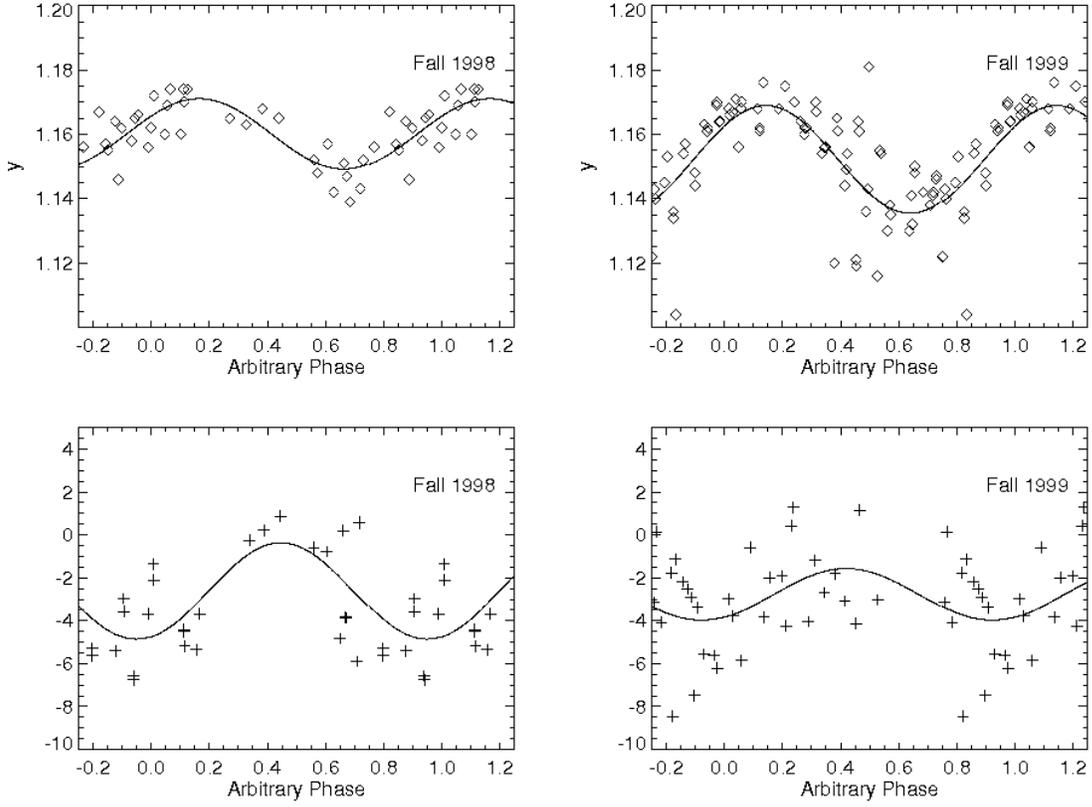} 
   \caption{\label{fig9} Plots of suspected radial pulsations seen in these data. On the top is the $y$ band Str\"{o}mgren photometry, and on the bottom are corresponding radial velocities. The periods are 17.8 days for the fall of 1998, and 13.4 days for the fall of 1999. Standard errors are less than 0.005 magnitudes for $y$, and about 1 km s$^{-1}$ for the radial velocities. $uvb$ magnitudes were also fitted, and the parameters of these fits are given in Table 4.}
\end{figure}

\placefigure{fig9}

\section{Discussion}

We have searched for high-velocity absorptions in a sample of B8Ia - A2 Ia supergiants that have been monitored at Ritter Observatory since 1992. Our initial results, which are shown in Table 5, begin to constrain the range of effective temperatures in which supergiant stars exhibit high-velocity absorption events. With the five years of data we have analyzed combined with the data presented in Kaufer et al.~(1996), only one high velocity absorption event has been observed in Deneb. These events are thus very rare for this star. Other stars of similar properties ($\nu$ Cep\footnote{
For $\nu$ Cep, there are numerous cases of enhanced absorption in the blue wing with maximum absorption around -100 km s$^{-1}$ and edge velocity around -200 km s$^{-1}$ as in the Kaufer et al.~(1996) definition, but the equivalent width involved is small. These velocities are sometimes hard to estimate because of water line contamination and/or indistinct edges. From the Mg II profiles in the Verdugo et al.~(1999) atlas, the terminal wind speed
of $\nu$ Cep is lower than Deneb's by about 35 km s$^{-1}$, so events will be harder to distinguish for that reason.} 
and 6 Cas) have not shown any signs of high velocity absorption in a long time series of observations. We can therefore conclude that Deneb is at the lower end of the range of effective temperature and luminosity ($T_{\rm eff}\sim$8500$\pm$200 K, $\log({L\over L_{\odot}})\sim$5.2; Aufdenberg et al.~2002) for which supergiants can exhibit high velocity absorption events.

\placetable{table5}

As discussed in Section 4.6, we found evidence for radial pulsations in the atmosphere of Deneb at two epochs during our five year investigation on this star. These pulsations were present for a few cycles and then vanished. Kaufer et al.~(1997) did not find radial pulsations for Deneb, although their data may have suggested such variability. Simultaneous photometry of the star is necessary for finding the radial pulsations because the atmosphere is also unstable to many non-radial modes simultaneously. The light curves at the epochs where we found radial pulsations have an amplitude such that they cannot be easily explained with non-radial modes (see e.g. Unno et al.~1979 for a description of non-radial oscillations). 

\section{Conclusions and Future Work}

Evidence was found for radial pulsations at two epochs during this campaign. Radial pulsations are somewhat rare in the atmosphere of Deneb, where the atmosphere is does not usally display these variations. With only two short epochs having radial pulsations, we find that Deneb displays radial pulsations $\le$20\% of the time. While photometry was difficult to obtain for such a bright target, it is necessary to find such behavior. A complete look at all the available data collected for this star will potentially lead to a better understanding of its interior. Gautschy (2009) has demonstrated that the variability of Deneb may be indicative of an underlying convective region for this star. Detailed analyses of the variability of a larger population of supergiants could demonstrate if these intermittent radial pulsations are normal for the population. 

A 40~day period was found with absorption events in 1997 as well as in the high velocity absorption event of 2001. A similar period is reported by Kaufer et al.~(1996) for the 1991 observations (H$\alpha$) of Deneb. Aufdenberg et al.~(2002) derived a stellar radius of 180 $R_{\odot}$ and $v \sin{i} = \approx 25$ km s$^{-1}$ for Deneb, which corresponds to a rotational period of 58 days. The $v \sin{i}$ was used in the spectral line fitting of the star, and no error bars were given, but a small increase of $\approx 10$ km s$^{-1}$ to this parameter will yield a rotational period of $\approx 40$ days. Similarly, the radius of the star could be reduced by $\approx 30$\% to yield this period. With a poorly determined parallax ($\pi = 2.31 \pm 0.32$ mas; van Leeuwen 2007), the interferometric radius determined by Aufdenberg et al.~(2002) cannot be used to constrain this parameter. Therefore, we conclude that the 40 day period is consistent with the rotation period of the star and these observations may indicate occasional coupling between the photosphere and the wind. 

During each observing season, the behavior of the H$\alpha$ profile was different. Therefore, long-term monitoring (over multiple seasons) is crucial to understanding the variability mechanisms of stars such as Deneb. With the extreme variability seen at the time of the high velocity absorption event in 2001, caution should be used on assigning mass-loss rates and other parameters to a single-epoch observation for Deneb and similar stars. Further work will be needed for the wind-momentum luminosity relationship (Kudritzki et al.~1999a,b) with the variability of these stars taken into account.

There is a need for observing campaigns at high spectral resolution for stars in the temperature and luminosity regime of Deneb. The high velocity absorption event we observed in 2001 is the only observed event for Deneb. High-resolution spectroscopic monitoring of Deneb at Ritter Observatory spans nearly two decades (1993 through present), and analysis of the entire data set is planned and will provide additional insights into the frequency of occurrence of high-velocity absorption events, of episodes of radial pulsation, and perhaps of other phenomena. Similar stars (in temperature and luminosity) need monitoring to see if Deneb is unique and where the temperature and luminosity boundary for high velocity absorption events is located.

\acknowledgments 
We acknowledge the anonymous referee for making suggestions that improved this paper. We are grateful for the Ritter Observing team who performed the spectroscopic observations during the calendar years 1997 through 2001, especially Karen Bjorkman, Anatoly Miroshnichenko, David Knauth, Chris Mulliss, Howard Ritter, Tracy Smith, Will Fischer, Patrick Ouellette, John Wisniewski, and Josh Thomas. The spectra were reduced by Chris Mulliss, David Knauth, and John Wisniewski by means of a script written by R. C. Dempsey, and later modified by J. P. Aufdenberg. Chris Mulliss created the first dynamical spectrum of Deneb from Ritter spectra, and began this project. Amber Ferguson analyzed radial velocity standard stars which helped to understand the error associated with our measurements; she was supported by NSF-PREST. We would like to thank Observatory Technician Robert Burmeister for keeping the telescope in excellent working condition. This research has made use of the SIMBAD database, operated at CDS, Strasbourg, France. Support for E. Kryukova on this project came from the NSF--REU program. Support for Ritter Observatory during the time of the observations was provided by the Fund for Astrophysical Research and the American Astronomical Society Small Grants Program. Support for FCAPT during the time period came from NSF grant AST-0587381. Special thanks to Louis J. Boyd, George McCook, and Robert J. Dukes, Jr. for their help in keeping the FCAPT in good working order. We thank Douglas Gies (Georgia State University) for helpful discussions related to the pulsational behavior of Deneb and for some software used in Figure 7. 

{\it Facilities:} \facility{Ritter Observatory,} \facility{Four College Automated Photoelectric Telescope}

\clearpage

\clearpage

\clearpage

\clearpage

\clearpage

\clearpage

\clearpage

\clearpage
 
\clearpage


\clearpage
\begin{deluxetable}{ccccccc}
\tablecaption{Spectroscopic Measurements of the H$\alpha$ profile and Si II doublet. This table can be found in its entirety online.}
\tablewidth{0pt}
\tablehead{
  \colhead{HJD}         &
  \colhead{$W_{\lambda}$}  &
  \colhead{$W_{\lambda}$}  &  
  \colhead{$W_{\lambda}$}  &
  \colhead{$V_r$} &
  \colhead{$V_r$} &
  \colhead{$\sigma(V_r)$} \\  
  \colhead{}         &
  \colhead{H$\alpha$, net}  &  
  \colhead{emission}         &  
  \colhead{absorption}         &  
  \colhead{Si II 6347}         &
  \colhead{Si II 6371}  & 
  \colhead{}  \\ 
  \colhead{}  &
  \colhead{(\AA )}  &
  \colhead{(\AA )} &
  \colhead{(\AA )} &
  \colhead{(km s$^{-1}$)} &
  \colhead{(km s$^{-1}$)} &
  \colhead{(km s$^{-1}$)} }    
\startdata
2450546.91051251  &  1.46  &  -0.29  &  1.75  &  -5.18  &  -4.88  &  0.81  \\
2450588.86303180  &  1.24  &  -0.21  &  1.45  &  -1.50  &  -1.91  &  0.85  \\
2450590.84395394  &  1.26  &  -0.21  &  1.48  &  -1.75  &  -1.57  &  0.77  \\
2450591.85544913  &  1.28  &  -0.18  &  1.46  &  -1.73  &  -1.59  &  0.76  \\
2450604.79591576  &  1.21  &  -0.20  &  1.42  &  -5.35  &  -4.69  &  1.00  \\
2450609.87109713  &  1.26  &  -0.20  &  1.46  &  -4.20  &  -4.19  &  0.75  \\
2450614.84276841  &  1.37  &  -0.22  &  1.59  &  -4.43  &  -4.53  &  0.76  \\
2450623.77241360  &  1.13  &  -0.27  &  1.40  &  -2.05  &  -1.91  &  0.76  \\
\enddata
\end{deluxetable}
\begin{deluxetable}{ccccccccc}
\tablecaption{Str\"omgren Photometry of Deneb. This table can be found in its entirety online.}
\tablewidth{0pt}
\tablehead{
  \colhead{HJD}         &
  \colhead{$u$}  &
  \colhead{$v$}  &  
  \colhead{$b$}  &
  \colhead{$y$} &
  \colhead{$u-b$} &
  \colhead{$b-y$} &
  \colhead{$c_1$} &
  \colhead{$m_1$} }    
\startdata
2450750.6404 & 2.229 & 1.347 & 1.172 & 1.171 & 1.057 & 0.001 & 0.707 & 0.174 \\ 
2450751.6399 & 2.222 & 1.336 & 1.164 & 1.158 & 1.058 & 0.006 & 0.714 & 0.166 \\ 
2450752.6566 & 2.215 & 1.332 & 1.158 & 1.153 & 1.057 & 0.005 & 0.709 & 0.169 \\ 
2450753.6364 & 2.210 & 1.322 & 1.155 & 1.141 & 1.055 & 0.014 & 0.721 & 0.153 \\ 
2450754.6564 & 2.197 & 1.314 & 1.142 & 1.142 & 1.055 & 0.000 & 0.711 & 0.172 \\ 
2450756.6545 & 2.174 & 1.299 & 1.124 & 1.121 & 1.050 & 0.003 & 0.700 & 0.172 \\ 
2450757.6456 & 2.178 & 1.295 & 1.122 & 1.119 & 1.056 & 0.003 & 0.710 & 0.170 \\ 
2450760.6463 & 2.181 & 1.312 & 1.134 & 1.128 & 1.047 & 0.006 & 0.691 & 0.172 \\
\enddata
\end{deluxetable}

\begin{deluxetable}{ccc}
\tablecaption{\label{table3} Str\"{o}mgren Photometric Parameters derived from the data.}
\tablewidth{0pt}
\tablehead{
  \colhead{Filter}         &
  \colhead{Magnitude} &
  \colhead{Amplitude of}\\
  \colhead{or Color}	&
  \colhead{}	&
  \colhead{Variability} }    
\startdata
$u$	&	2.202	& 	0.018 \\
$v$	&	1.331	&	0.016 \\
$b$	&	1.152	&	0.014 \\
$y$	&	1.149	&	0.014 \\
$u-b$	&	1.050	&	0.009 \\
$b-y$	&	0.003	&	0.005 \\
$c_1$	&	0.692	&	0.017 \\
$m_1$	&	0.177	&	0.010 \\
\enddata
\end{deluxetable}

\begin{deluxetable}{cccccc}
\tablecaption{\label{table4} Coefficients for the sine wave fits (through $\chi^2$ minimization) to the radial pulsations shown in Figure 9 as well as other variables we fitted. The errors are the standard errors of the fit.}
\tablewidth{0pt}
\tablehead{
  \colhead{Season}  &
  \colhead{Type}  &
  \colhead{Period}  &
  \colhead{$A$}         &
  \colhead{$x_c$} &
  \colhead{$c$}	\\
  \colhead{}  &
  \colhead{}  &
  \colhead{(days)}  &
  \colhead{(km s$^{-1}$) or (mag)}         &
  \colhead{(phase)} &
  \colhead{(km s$^{-1}$) or (mag)}} 
\startdata
Fall 1998	&	$u$ 	&	17.8	&	$-0.016\pm0.002$	&	$0.39\pm0.02$	&	$2.208\pm0.001$	\\
Fall 1998	&	$v$ 	&	17.8	&	$-0.011\pm0.002$	&	$0.42\pm0.04$	&	$1.335\pm0.002$	\\
Fall 1998	&	$b$ 	&	17.8	&	$-0.011\pm0.002$	&	$0.38\pm0.04$	&	$1.162\pm0.002$	\\
Fall 1998	&	$y$ 	&	17.8	&	$-0.011\pm0.002$	&	$0.42\pm0.03$	&	$1.160\pm0.001$	\\
Fall 1998	&	Si II RV &	17.8	&	$2.27\pm0.62$		&	$0.19\pm0.03$	&	$-2.56\pm0.43$	\\ \hline
Fall 1999	&	$u$ 	&	13.4	&	$-0.023\pm0.002$	&	$0.38\pm0.02$	&	$2.211\pm0.002$	\\
Fall 1999	&	$v$ 	&	13.4	&	$-0.016\pm0.002$	&	$0.39\pm0.02$	&	$1.336\pm0.001$	\\
Fall 1999	&	$b$ 	&	13.4	&	$-0.017\pm0.002$	&	$0.39\pm0.02$	&	$1.155\pm0.001$	\\
Fall 1999	&	$y$ 	&	13.4	&	$-0.017\pm0.002$	&	$0.39\pm0.02$	&	$1.152\pm0.001$	\\
Fall 1999	&	Si II RV &	13.4	&	$1.21\pm0.57$		&	$0.17\pm0.08$	&	$-2.76\pm0.42$  \\ \hline
\enddata
\end{deluxetable}

\begin{deluxetable}{ccccccccc}
\tablecaption{\label{table5} Hot supergiant stars observed at Ritter Observatory with the same spectroscopic instrumentation as the observations presented here. All stars with an absence of the absorption events have more than 40 observations. Presented here are fundamental properties, the number of observations, and whether high velocity absorption events have been seen in our data. Kaufer et al.~(1996) also found high velocity absorption in the late B-type supergiants HD 91619 and HD 96919, which both are hotter and more luminous than Deneb and are inaccessible to the instrumentation used in this survey. $\nu$ Cep is nearly identical to Deneb, but has not exhibited any high velocity absorption events.}
\tablewidth{0pt}
\rotate
\tablehead{
  \colhead{Star}         &
  \colhead{MK}  &
  \colhead{$T_{\tt eff}$}  &  
  \colhead{$\log (L/L_{\odot})$} &
  \colhead{Reference} &
  \colhead{Number} &
  \colhead{Years} &
  \colhead{Absorption}\\
    \colhead{} &
    \colhead{} &
    \colhead{K} &
    \colhead{} &
    \colhead{(Stellar Parameters)} &
    \colhead{of Spectra} &
    \colhead{of Data} &
    \colhead{Events?} }    
\startdata
Rigel	&     	B8 Ia	&   	12500	&  	5.56 	&      	Markova et al.~(2008)		& 203	& 1996-2007	& yes \\
HR 8020	&       B8Iae   &	13000 	&	5.08    &	Markova et al.~(2008)		& 29	& 2000-2006	& yes \\
HR 1035	&       B9 Ia   &	9730    &	4.92    &	Verdugo et al.~(1999)		& 108	& 1994-2007	& yes \\
HR 1040	&      	A0 Ia	&	9600    &	4.76    &  	Barlow \& Cohen (1977) 		& 187	& 1993-2007	& yes \\
9 Per	&      	A2 Ia   &	9000	&	4.99    &	Kudritzki et al.~(1999a)	& 40	& 1993-2005	& yes \\
Deneb 	&       A2 Ia   &	8500	&	5.20  	&	Aufdenberg et al.~(2002)	& 781 	& 1993-2007	& yes, rare \\ \hline
$\nu$ Cep &    	A2 Ia  	&	8500	&	4.76	&					& 119	& 1994-2007	& no, See Footnote 3  \\ \hline
6 Cas 	&     	A3 Ia  	&	8400	&	5.13	&	Kudritzki et al.~(1999)		& 69	& 1993-2007	& no  \\
\enddata
\end{deluxetable}
\end{document}